\documentclass{article}

    \PassOptionsToPackage{numbers,square}{natbib}
 \usepackage[preprint]{neurips_2026}

\usepackage[utf8]{inputenc} 
\usepackage[T1]{fontenc}    
\usepackage{hyperref}       
\usepackage{url}            
\usepackage{booktabs}       
\usepackage{amsfonts}       
\usepackage{nicefrac}       
\usepackage{microtype}      
\usepackage{xcolor}         

\usepackage{graphicx}
\usepackage[dvipsnames]{xcolor}
\usepackage{booktabs}
\usepackage{mathrsfs}
\usepackage{xspace}
\usepackage{amsthm}
\usepackage{amsmath,amssymb,amsfonts}
\usepackage[ruled, vlined]{algorithm2e}
\usepackage{algpseudocode}
\usepackage{array}
\usepackage{multirow}
\usepackage{fontawesome5}
\usepackage{makecell}
\usepackage{textcomp}
\usepackage{url}
\usepackage{verbatim}
\usepackage{subcaption}
\usepackage{tcolorbox}
\usepackage{enumitem}
\usepackage{wrapfig}

\newcommand{\ourmethod}{\textsc{ATOM}\xspace}
\newcommand{\down}[1]{$^{\,\color{BlueGreen}\downarrow #1}$}
\newcommand{\up}[1]{$^{\,\color{RedOrange}\uparrow #1}$}

\title{\ourmethod: Instantiating Budget-Controllable Multi-Agent Collaboration via Nucleus-Electron Hierarchy}

\author{%
  \textbf{Xinkui Zhao}$^{1,2,3}$,
  \textbf{Sai Liu}$^{1}$,
  \textbf{Yifan Zhang}$^{1}$\textsuperscript{\thanks{Corresponding author.}},
  \textbf{Qingyu Ma}$^{1}$,
  \textbf{Zewen Lin}$^{1}$,\\
  \textbf{Naibo Wang}$^{2,1,3}$,
  \textbf{Guanjie Cheng}$^{2,1,3}$,
  \textbf{Chang Liu}$^{2,1,3}$,
  \textbf{Yueshen Xu}$^{4}$\\
\small $^{1}$Zhejiang University, $^{2}$Ningbo Global Innovation Center, Zhejiang University \\
\small $^{3}$Zhejiang Key Laboratory of Digital-Intelligence Service Technology, $^{4}$Xidian University \\
\tt \small \{zhaoxinkui, liusai2024, 12451018, maqingyu, 22551341\}@zju.edu.cn,\\ 
\tt \small \{wangnaibo, chengguanjie, chang.liu\}@zju.edu.cn, ysxu@xidian.edu.cn
}

\begin{document}

\maketitle

\begin{abstract}
  Large Language Model (LLM)-based multi-agent systems rely on optimized collaboration topologies to balance performance and communication costs. However, current methods struggle with the inherent stability-extensibility trade-off and often misalign computational budgets with query difficulty. We propose \textsc{ATOM}, an adaptive framework that generates budget-controllable collaboration graphs via a novel task-driven reinforcement learning paradigm. Inspired by atomic structures, \textsc{ATOM} employs a nucleus-electron hierarchy: it maintains a stable, offline-learned collaboration backbone (the nucleus) while dynamically activating query-conditioned agents (electrons) during inference. Crucially, a complexity-aware budgeting strategy aligns resource consumption with task demands by estimating query difficulty to strictly regulate electron instantiation. Extensive experiments across six diverse benchmarks demonstrate that \textsc{ATOM} achieves state-of-the-art performance while improving token efficiency by up to $30\%$ compared to strong baselines.
\end{abstract}

\section{Introduction}

\begin{wrapfigure}[18]{r}{0.55\textwidth}
  \centering
  \begin{minipage}[t]{\linewidth}
    \centering
    \includegraphics[width=\linewidth]{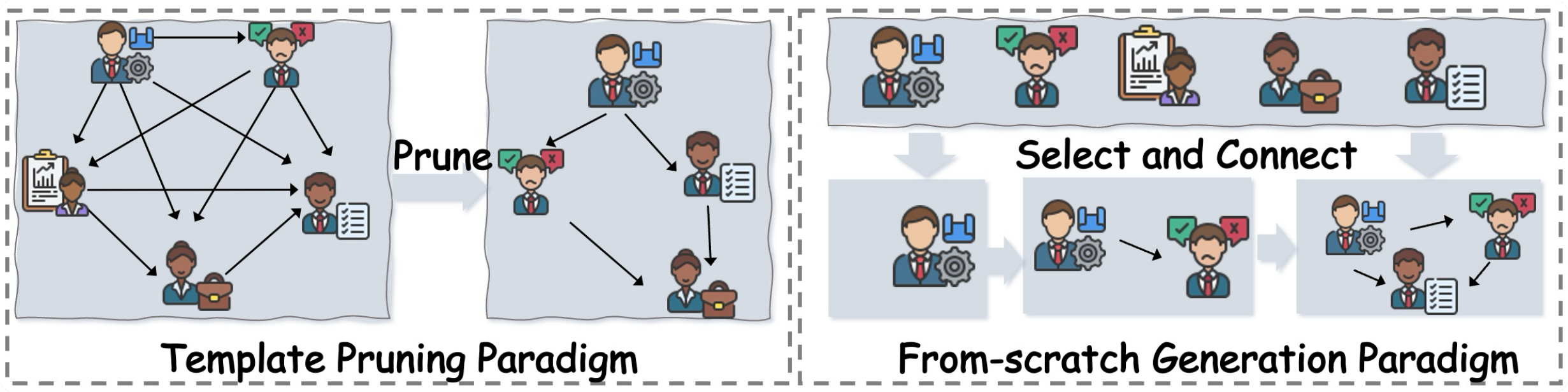}
    \caption*{(a) Existing MAS topology design methods.}
  \end{minipage}

  \begin{minipage}[t]{\linewidth}
    \centering
    \includegraphics[width=\linewidth]{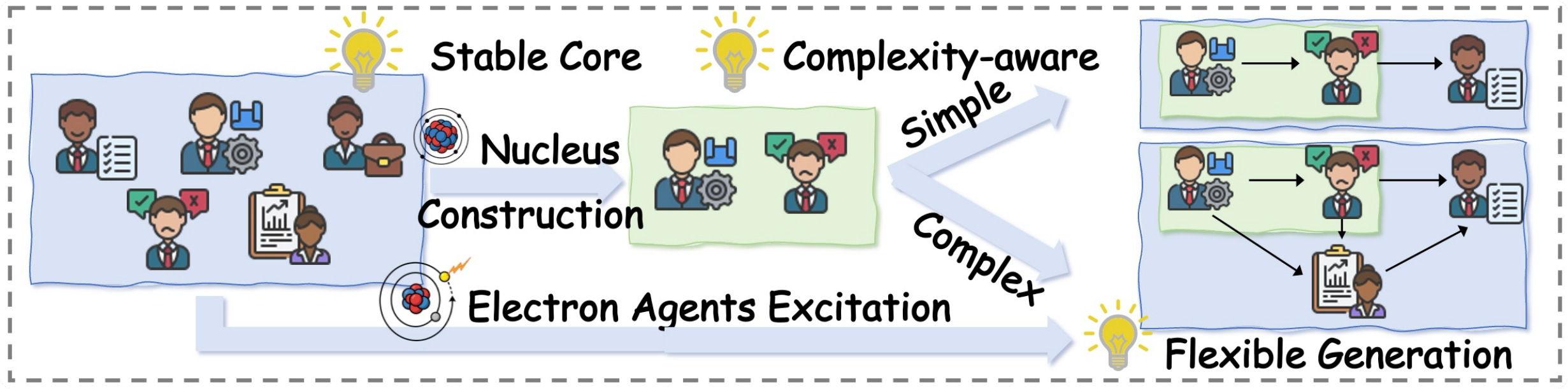}
    \caption*{(b) \ourmethod topology optimization paradigm.}
  \end{minipage}
  \caption{Comparison of MAS topology design paradigms.}
  \label{fig:topology_paradigms}
\end{wrapfigure}

Large language model (LLM)-based agents show strong capabilities across diverse domains~\citep{liu2024large,zhong2024debug,dong2024self,xie2024waitgpt,li2024dawn,hong2025data,zhao2025video,yue2025survey,zhu2024autotqa}, yet single agents struggle with complex problems due to limited expertise and reasoning depth~\citep{du2023improving,valmeekam2023planning,ke2025survey}. This has driven the shift toward multi-agent systems (MAS), which leverage collective intelligence through specialized roles~\citep{islam2024mapcoder,zhang2025sortinghat}. Crucially, MAS effectiveness relies heavily on its \textbf{collaboration topology}---the graph dictating agent organization and information routing~\citep{zhang2024aflow,zhou2025multi,wei2022chain,liu2025graph,zhuge2024gptswarm}. Consequently, designing task-adaptive collaboration graphs is a central research challenge.

Previous design methodologies for orchestrating multi-agent systems rely heavily on static, manual structures~\citep{wei2022chain,hong2023metagpt,yao2023tree}. While these hand-crafted templates improve system capabilities and ensure a reliable execution flow, they are inherently rigid. Their inability to dynamically adapt to varying task difficulties often leads to suboptimal reasoning pathways and inefficient resource allocation.

To overcome the limitations of fixed architectures, researchers have recently explored dynamic topology generation methods~\citep{zhuge2024gptswarm,shen2025understanding}. Prominent directions include template modification~\citep{zhang2024cut,wang2025agentdropout} and generative graph modeling~\citep{zhang2024g,li2025assemble}. 
Although these dynamic approaches show promising progress, scaling them for practical deployment exposes significant challenges. Specifically, constructing collaboration topologies that are simultaneously effective, efficient, and executable remains constrained by two universal limitations: \textbf{1) Scalability and Efficiency:} Navigating the combinatorial connection space is a major hurdle. Modification methods rely on bloated, fully-connected initial super-graphs, incurring massive upfront communication overhead. Conversely, purely generative methods attempt to predict or decode routing pathways entirely dynamically, forcing the system to navigate an exponentially growing search space. \textbf{2) Budget Misalignment:} Current approaches decouple topological scale from query complexity. Without explicit difficulty estimation, modification methods over-spend tokens on trivial queries before simplification occurs. Similarly, generative methods fall into an ``average-complexity'' trap---over-provisioning resources for simple tasks while under-provisioning for complex ones.

To address these gaps, we propose \ourmethod, an \textbf{A}daptive \textbf{T}opology \textbf{O}ptimization \textbf{M}echanism using task-driven reinforcement learning. \ourmethod achieves robustness and adaptability by structuring the MAS as an ``atom''. It maintains a stable \emph{nucleus} for reusability while dynamically activating \emph{electrons} to expand communication on demand, via two key features:

\noindent\textbf{(1) Nucleus--Electron Two-Stage Topology Learning.} To prevent structural volatility and mitigate the topology search overhead of from-scratch generation, we decouple topology learning into an offline and online phase. We first condense a stable \emph{nucleus} (core backbone) offline to serve as a resilient reasoning conduit. During inference, the online generator only predicts dual-channel edge probabilities (spatial and temporal) for query-conditioned \emph{electrons}, drastically reducing the search space while ensuring structural control.

\noindent\textbf{(2) Complexity-Aware Budgeted Instantiation.} To break the ``average-complexity'' trap inherent in static or generative baselines, we explicitly align computational resources with task demands. We introduce a predictor to estimate query difficulty, which dynamically allocates a adaptive collaboration budget. This ensures true token efficiency by instantiating a sparse electron subgraph strictly when the task complexity warrants it, preventing resource waste on trivial queries.

Our main contributions are summarized as follows:

\begin{itemize}[leftmargin=*]
    \item \textbf{Hierarchical Nucleus-Electron Architecture.} We propose a novel structural paradigm that decouples offline stable backbone condensation (nucleus) from online task-driven expansion (electrons). This inherently resolves the stability-extensibility trade-off while significantly mitigating the topology search overhead of from-scratch generation.

    \item \textbf{Complexity-Aware Budgeted Instantiation.} We introduce a difficulty estimator that dynamically allocates multi-agent computation and regulates electron subgraph instantiation based on query complexity. To optimize this budget-constrained routing policy, we design a task-driven reinforcement learning objective with structural regularization.

    \item \textbf{State-of-the-Art Empirical Performance.} Extensive experiments across six diverse benchmarks demonstrate the superiority of \ourmethod. It consistently achieves state-of-the-art accuracy while improving token efficiency by up to 30\% compared to strong baselines, establishing a new Pareto frontier for MAS collaboration.
\end{itemize}

\section{Problem Formulation}
\label{sec:problem_formulation}

\subsection{Multi-Agent System as a Spatial-Temporal Graph}
To formalize the budget-constrained adaptive topology design problem, we conceptualize the MAS as a dynamic spatial-temporal graph. Consider a MAS comprising a global role pool $\mathcal{V}$ operating over $T$ conversational rounds. For a query $q$, we formulate the multi-round orchestration as a dynamic information routing process over a composite directed graph $\mathcal{G}_{q} = (\mathcal{V}_{q}, \mathcal{E}_{q}^{(S)}, \mathcal{E}_{q}^{(T)})$, where $\mathcal{V}_{q} \subseteq \mathcal{V}$ denotes the instantiated subset of active agents. Each agent $v_i \in \mathcal{V}_{q}$ is associated with a specific \textit{functional role} $\rho_i$ and a \textit{contextual state} $\sigma_i$. The topology is governed by two orthogonal edge sets:

\begin{itemize}[leftmargin=*]
    \item \textbf{Spatial Edges} ($\mathcal{E}_{q}^{(S)}$): Define within-round synchronous communication. An edge $(u, v)$ indicates agent $v$ aggregates the immediate response from $u$. The spatial subgraph strictly forms a Directed Acyclic Graph (DAG) to prevent logical deadlocks.
    \item \textbf{Temporal Edges} ($\mathcal{E}_{q}^{(T)}$): Facilitate cross-round asynchronous memory retrieval, enabling agent $u$ at round $t$ to query historical context generated by $v$ in rounds $t' < t$.
\end{itemize}

\textbf{Execution Protocol.} The execution follows a topologically-ordered activation sequence. At round $t$, an active agent $v_i$ generates a message $m_i^{(t)}$ using its associated LLM, conditioned on its role, state, the query $q$, and aggregated messages from topological dependencies:

\begin{equation}
    m_i^{(t)} = \text{LLM}_i \Big( \text{Prompt} \big( \rho_i, \sigma_i, q, \{m_j^{(t)} \mid (v_j, v_i) \in \mathcal{E}_{q}^{(S)}\}, \{m_j^{(t')} \mid (v_j, v_i) \in \mathcal{E}_{q}^{(T)}\} \big) \Big)
    \label{eq:execution}
\end{equation}

\subsection{Budget-Constrained Topology Optimization}
\begin{figure}[t]
    \centering
    \begin{subfigure}[b]{0.32\textwidth}
        \centering
        \includegraphics[width=\textwidth]{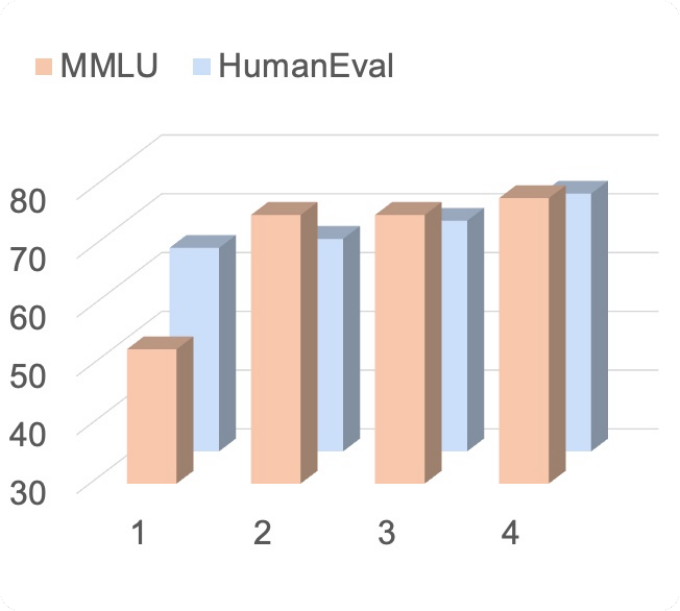}
        \caption{Easy Level}
        \label{fig:easy}
    \end{subfigure}
    \hfill
    \begin{subfigure}[b]{0.32\textwidth}
        \centering
        \includegraphics[width=\textwidth]{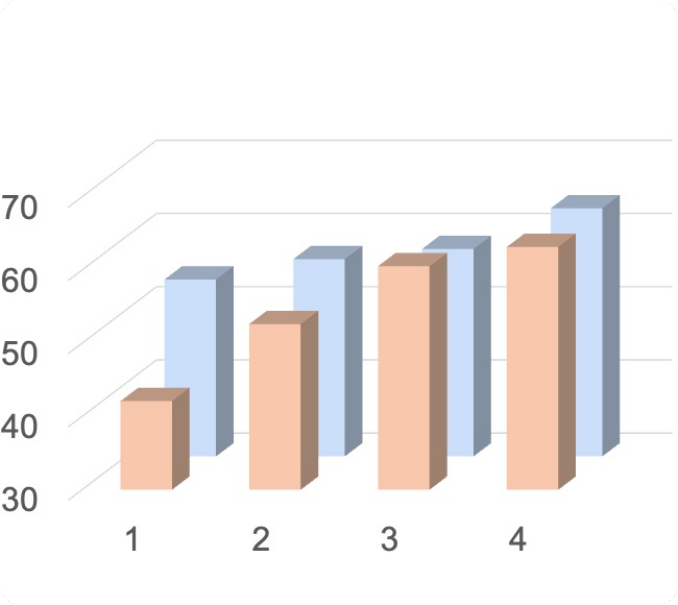}
        \caption{Medium Level}
        \label{fig:medium}
    \end{subfigure}
    \hfill 
    \begin{subfigure}[b]{0.32\textwidth}
        \centering
        \includegraphics[width=\textwidth]{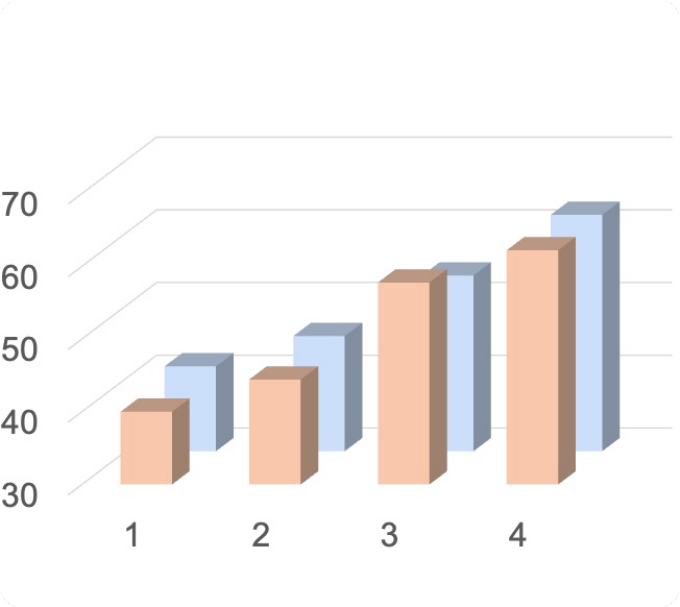}
        \caption{Hard Level}
        \label{fig:hard}
    \end{subfigure}
    
    \caption{Performance under different agent budgets across difficulty levels.}
    \label{fig:complexity_comparison}
\end{figure}

Our empirical analysis (Figure~\ref{fig:complexity_comparison}) reveals an ``average-complexity trap'' in MAS scaling. By stratifying tasks from our custom complexity-annotated meta-dataset 
into \emph{Easy}, \emph{Medium}, and \emph{Hard} tiers, we observe a clear divergence in resource utilization. For Easy queries, a minimal number of agents is sufficient to saturate performance, and scaling up the agent budget yields diminishing returns while substantially inflating communication costs. Conversely, Hard queries consistently benefit from a larger, scaled-up agent budget. Static, densely connected topologies are therefore fundamentally inefficient: they blindly overuse resources on simple queries while lacking sufficient flexibility to expand reasoning capacity for complex ones.

To resolve this tension, the system's topological scale must dynamically adapt to the query's intrinsic difficulty, since each agent instantiation and message-passing edge directly multiplies the operational cost. 

Therefore, we formalize the topology orchestration as a budget-constrained optimization problem. For a query $q \sim \mathcal{D}$, we learn a routing policy $\pi_{\theta}(\mathcal{G} \mid q)$ to maximize the expected task utility:
\begin{equation}
    \max_{\theta} \quad \mathbb{E}_{q \sim \mathcal{D}, \mathcal{G}_{q} \sim \pi_{\theta}} \left[ \mathcal{U}(\mathcal{G}_{q} \mid q) \right]
    \quad \text{s.t.} \quad
    \mathcal{C}(\mathcal{G}_{q}) \le \mathcal{B}(q),
    \label{eq:objective}
\end{equation}
where $\mathcal{U}(\cdot)$ denotes the task success utility, $\mathcal{C}(\cdot)$ represents the total token and computational cost of executing the topology $\mathcal{G}_{q}$, and $\mathcal{B}(q)$ is the query-dependent budget estimated by our complexity predictor (detailed in Section~3.2.1). This constraint strictly enforces that the instantiated MAS never exceeds the difficulty-aligned resource cap.

\section{\ourmethod for MAS Topology Design}
\label{sec:method}

\begin{figure}[t]
    \centering
    \includegraphics[width=1\linewidth]{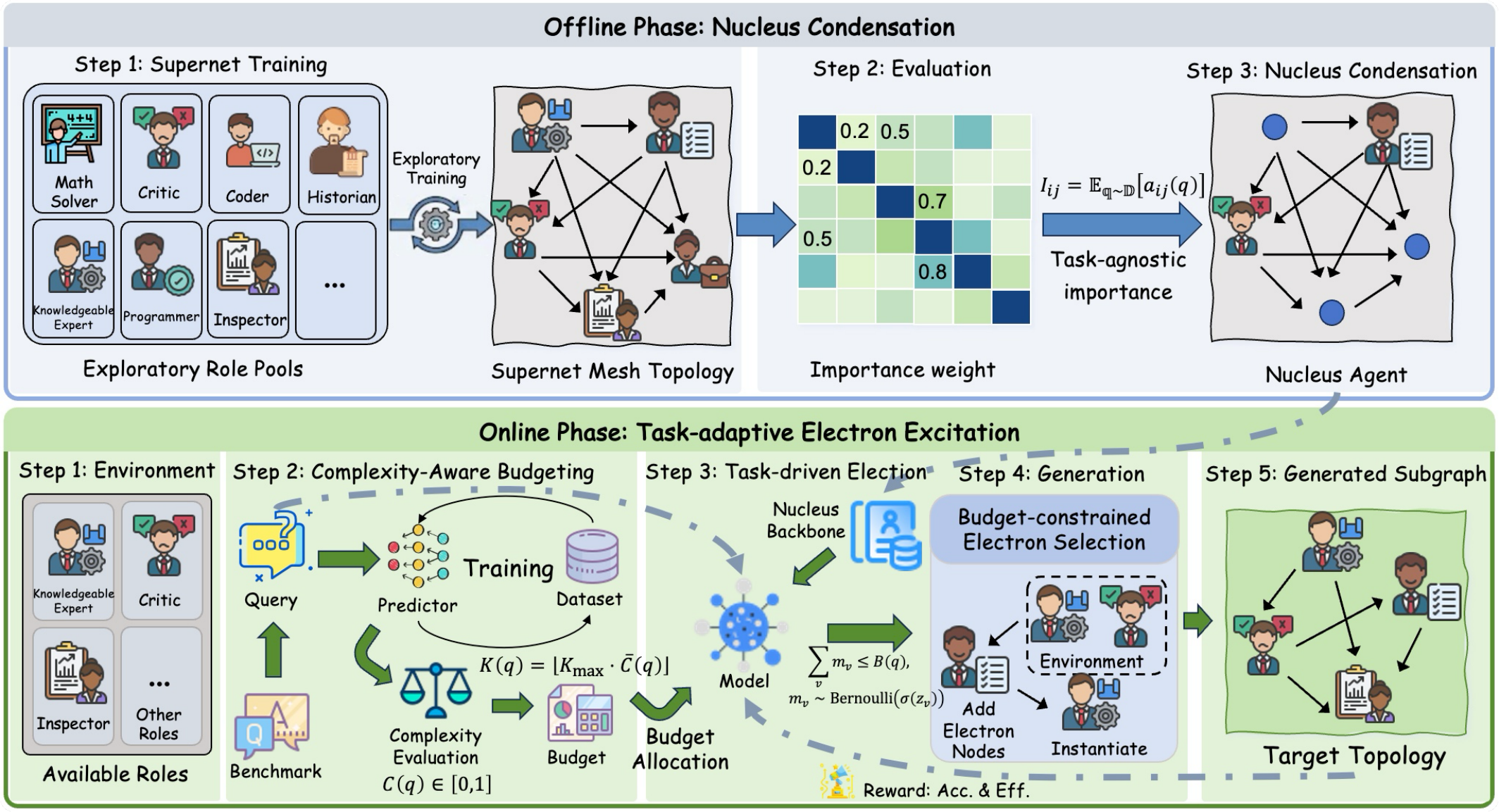}
    \caption{Pipeline of \ourmethod topology generation.}
    \label{fig:pipeline}
\end{figure}

To overcome the inherent stability-extensibility trade-off in existing architectures, \ourmethod employs a two-tier \emph{nucleus--electron} hierarchy. Specifically, we partition the global agent pool into a persistent, offline-learned \emph{nucleus} backbone ($\mathcal{V}_{\text{nuc}}$) and a dynamic \emph{electron} reservoir ($\mathcal{V}_{\text{elec}}$). Crucially, this backbone refers strictly to a fixed partition of \emph{agent nodes}; all inter-agent communication edges---including those within the nucleus itself---remain fully dynamic and query-conditioned during the online phase to maintain maximum reasoning flexibility.

\subsection{Offline Phase: Nucleus Backbone Construction}
\label{sec:offline_phase}

Prior to inference, \ourmethod learns a compact, domain-specific nucleus backbone to constrain the combinatorial search space of dynamic graph generation. To mitigate out-of-distribution brittleness, we condense a tailored nucleus for each reasoning domain using a dedicated training split.

\textbf{Stochastic Edge Parameterization.}
To optimize structural layouts over the discrete graph space, we define a stochastic routing policy over a fully connected supernet. Let $\theta_{i,j}^{(S)}$ and $\theta_{i,j}^{(T)}$ be the learnable structural logits for spatial and temporal edges. The edge activation probability is $\rho_{i,j} = \sigma(\theta_{i,j})$. We optimize these logits to maximize expected task utility $\mathcal{U} \in [0, 1]$ while enforcing structural compactness via a sparsity prior:
\begin{equation}
    \max_{\theta}\; \mathbb{E}_{\mathcal{G}\sim\pi_\theta}\big[\mathcal{U}(\mathcal{G})\big] - \lambda_{\text{sparse}} \sum_{k \in \{S, T\}} \frac{1}{|\mathcal{E}^{(k)}|} \sum_{(i,j)\in\mathcal{E}^{(k)}} \sigma(\theta_{i,j}^{(k)})
\end{equation}
We optimize this objective via the REINFORCE algorithm, ensuring reasoning pathways that consistently contribute to successful rollouts are rewarded, while redundant channels decay.

\textbf{Topological Condensation and Partitioning.}
Upon convergence, $\rho_{i,j}$ serves as a reliable proxy for an edge's marginal utility. Rather than learning isolated node parameters, deriving the nucleus from converged edge distributions ensures we capture the \emph{collaborative synergy} between agents. We evaluate each agent's foundational necessity using its Expected Degree Centrality (EDC). 
Given that temporal connections function primarily as passive historical retrievals rather than active collaborative pathways, we evaluate an agent's structural importance strictly over the spatial subgraph. Thus, the topological saliency $\mathcal{S}_{\text{node}}(v_i)$ is defined as:
\begin{equation}
    \mathcal{S}_{\text{node}}(v_i) = \sum_{j} (\rho_{i,j}^{(S)} + \rho_{j,i}^{(S)})
\end{equation}
Finally, the $K_{nuc}$ agents with the highest saliency constitute the stable nucleus ($\mathcal{V}_{\text{nuc}}$). 
This encourages a highly utilized core, relegating all remaining agents to the dynamic electron reservoir ($\mathcal{V}_{\text{elec}}$).

\subsection{Online Phase: Electron Excitation}
\label{sec:online_phase}

Building on the fixed nucleus partition, we formulate online electron activation as a reinforcement learning problem. To effectively regulate computational costs, we use the number of active agents as a deterministic proxy for token consumption, as bounding agent activation inherently bounds the maximum communication rounds and context window growth. Accordingly, the generative policy $\pi_\theta(\mathcal{G}\mid q)$ selects a query-conditioned topology to maximize expected task utility, constrained by a complexity-derived agent budget:
\begin{equation}
    \max_{\theta} \ \mathbb{E}_{q \sim \mathcal{D}, \mathcal{G}_q \sim \pi_\theta}
    \Big[ \mathcal{U}\big(\mathcal{G}_q \mid q\big) \Big]
    \quad \text{s.t.} \quad
    |\mathcal{V}_{\text{elec}}(q)| \leq K(q)
    \label{eq:objective2}
\end{equation}

\subsubsection{Complexity-Aware Budgeting}
\label{sec:tech_budget}
We introduce a budgeting mechanism that maps query $q$ to a normalized scalar $\bar{C}(q) \in [0,1]$, imposing an adaptive upper limit on the active electron pool. Since predicting precise algorithmic complexity from surface text is intractable, we use structural density as a proxy. We construct a composite feature vector $\mathbf{x}_q = [ \mathbf{W}_{\text{PCA}} f_{\text{enc}}(q) \,\|\, \Phi(q) ]$, where $\Phi(q)$ explicitly quantifies symbolic and logic densities (e.g., numerical variables, logical operators) that strongly correlate with the required combinatorial search space.

Because the intrinsic scale of difficulty varies across domains, a domain-gated mechanism is employed during inference. Given the dataset embedding $\mathbf{e}_d$, we generate a feature-wise gating vector $\mathbf{g}_d = \sigma(\text{MLP}_{\phi}(\mathbf{e}_d))$ to prevent domain confusion. The complexity score is then efficiently computed via the core regressor:
\begin{equation}
    \bar{C}(q) = \text{MLP}_{\theta}( \mathbf{x}_q \odot \mathbf{g}_d )
\end{equation}
This scalar dictates the active electron budget via $K(q) = \lfloor K_{\max} \cdot \bar{C}(q) \rfloor$, acting as an adaptive capacity ceiling that enables peripheral capacity only for constraint-heavy tasks.

\subsubsection{Dynamic Activation and Instantiation}
\label{sec:tech_instantiation}

To capture non-linear relational dependencies between any feature vectors $\mathbf{x}$ and $\mathbf{y}$, we define a unified dyadic interaction operator:
\begin{equation}
    \Psi_{\text{rel}}(\mathbf{x}, \mathbf{y}) = \Big[ \mathbf{x} \,\|\, \mathbf{y} \,\|\, \mathbf{x} \odot \mathbf{y} \,\|\, |\mathbf{x} - \mathbf{y}| \Big]
\end{equation}

\textbf{Query-Conditioned State Representation.}
To avoid the quadratic overhead of self-attention, we inject task semantics $\mathbf{f}_q \in \mathbb{R}^{d_0}$ into initial agent features $\mathbf{e}_i^{(0)}$ via a lightweight $O(N)$ row-wise gating mechanism:
\begin{equation}
    \mathbf{h}_i = \mathrm{LayerNorm}\!\left( (\mathbf{W}_{\mathrm{in}} \mathbf{e}_i^{(0)}) \odot \sigma(\mathbf{W}_{\mathrm{gate}} \mathbf{f}_q) + \mathbf{W}_{\mathrm{ctx}} \mathbf{f}_q \right)
\end{equation}

\textbf{Task-Node Compatibility \& Constrained Activation.}
The unnormalized activation logit $\alpha_i$ for candidate electron $v_i$ combines a task-conditioned relational score with a semantic alignment term:
\begin{equation}
    \alpha_i = \Phi_{\text{node}}\big(\Psi_{\text{rel}}(\mathbf{h}_i, \mathbf{W}_{\mathrm{ctx}} \mathbf{f}_q)\big) + \cos\!\big(\psi_K(\mathbf{e}_i^{(0)}), \psi_Q(\mathbf{f}_q)\big)
\end{equation}
We deliberately use the raw feature $\mathbf{e}_i^{(0)}$ in the cosine term to prevent redundant re-injection of query information. While nucleus agents remain permanently active, the electron indicator vector $\mathbf{y} \in \{0,1\}^{|\mathcal{V}_{\mathrm{elec}}|}$ is recruited via conditional Bernoulli sampling to strictly respect the predicted budget $K(q)$:
\begin{equation}
    P(\mathbf{y}\mid q) \propto \prod_i \sigma(\alpha_i)^{y_i}\bigl(1-\sigma(\alpha_i)\bigr)^{1-y_i} \cdot \mathbb{I}\!\Big(\textstyle\sum_i y_i \le K(q)\Big)
\end{equation}
To sample exactly from this constrained distribution without distorting probabilities via ad-hoc Top-$K$ truncation, we implement dynamic programming. Let the odds ratio be $w_i = \sigma(\alpha_i)/(1-\sigma(\alpha_i))$. The suffix partition function follows $Z_i^{(b)} = Z_{i+1}^{(b)} + w_i Z_{i+1}^{(b-1)}$. Crucially, to enforce the inequality constraint ($\le b$) rather than exact-$K$ matching, we initialize $Z_{N+1}^{(b)} = 1$ for $b \ge 0$, and $0$ otherwise. Given the dynamically updated remaining budget $r_i$, the exact conditional activation probability is $P(y_i=1 \mid y_{<i}, r_i) = w_i Z_{i+1}^{(r_i-1)} / Z_i^{(r_i)}$. 

\textbf{Relational Topology Synthesis.}
For active pairs, dyadic representations $\mathbf{z}_{i,j} = \Psi_{\text{rel}}(\mathbf{h}_i, \mathbf{h}_j)$ map to dual routing logits $\mathcal{L}_{i,j}^{(S)}$ and $\mathcal{L}_{i,j}^{(T)}$. Spatial edges are sampled sequentially under a strict acyclicity constraint to guarantee an executable DAG for intra-round message passing. Conversely, temporal edges are sampled independently across rounds without acyclicity constraints, enabling flexible historical memory retrieval.

\subsubsection{Task-Driven RL Training Strategy}
\label{sec:tech_rl}

We optimize the generative policy $\pi_\theta$ via a REINFORCE-based objective. The composite reward balances reasoning efficacy with topological parsimony: $\mathcal{R}(\mathcal{G}) = r_{\text{task}} - \lambda_{\text{cost}} (|\mathcal{E}_{\text{active}}|/|\mathcal{E}_{\text{max}}|)$. We compute a standardized advantage $\hat{A}$ using a moving baseline to constrain variance.

\textbf{Sequential DAG-Constrained Surrogate.}
During the DAG-constrained sampling of spatial edges, candidate connections that would create cycles are deterministically skipped. Consequently, we utilize a length-normalized sequential log-probability surrogate $\tilde{l}_{\text{edge}}$ to prevent large graphs from dominating gradient updates:
\begin{equation}
    \tilde{\ell}_{\text{edge}} = \frac{1}{T_S} \sum_{t=1}^{T_S} \log \pi_\theta(a_t \mid a_{<t}, \mathcal{V}_{\text{active}}, q) + \frac{1}{T_T} \sum_{t=1}^{T_T} \log \pi_\theta(b_t \mid \mathcal{V}_{\text{active}}, q)
\end{equation}
where $T_S$ and $T_T$ are the counts of actually evaluated spatial and temporal edges. 

\textbf{Structural Hierarchy Regularization.}
We maximize the comprehensive objective $\mathcal{J}(\theta) = \mathbb{E}_{\pi_\theta} [ ( \log \pi_\theta(\mathcal{V}_{\text{active}} \mid q) + \tilde{\ell}_{\text{edge}} ) \hat{A} ] - \gamma \mathcal{L}_{\text{struct}}$. To bias the routing policy toward nucleus-centered communication, $\mathcal{L}_{\text{struct}}$ is formulated as a dual-margin ranking loss:
\begin{equation}
    \mathcal{L}_{\text{struct}} = \sum_{k \in \{nn, ne\}} \max\Big(0, m - \big( \mathbb{E}_{e \in \mathcal{E}_{k}}[\rho_{e}] - \mathbb{E}_{e \in \mathcal{E}_{ee}}[\rho_{e}] \big) \Big)
\end{equation}
where $\mathcal{E}_{nn}$, $\mathcal{E}_{ne}$, and $\mathcal{E}_{ee}$ represent the candidate non-self-loop directed edges under the fixed node partition. The expectation is strictly computed over the dense pre-sampling routing probabilities ($\rho_e$). Crucially, in the absence of dense step-wise task rewards, this regularization provides a continuous behavioral signal that stabilizes early exploration, penalizing topologies in which peripheral electron-to-electron communication dominates nucleus-centered communication.

\begin{table*}[t]
\centering
\small
\renewcommand{\arraystretch}{1.3}
\setlength{\tabcolsep}{5pt}

\resizebox{\linewidth}{!}{
\begin{tabular}{l | cccccc | c}
\toprule
\textbf{Method} & \textbf{MMLU} & \textbf{GSM8K} & \textbf{HumanEval} & \textbf{AQuA} & \textbf{MultiArith} & \textbf{SVAMP} & \textbf{Average} \\
\midrule
Vanilla      & 54.90 & 78.85 & 65.29 & 57.01 & 92.68 & 84.39 & 72.27 \\ 
CoT          & 55.56 \up{0.66} & 81.17 \up{2.32} & 67.77 \up{2.48} & 54.21 \down{2.80} & 94.64 \up{1.96} & 84.69 \down{0.20} & 73.01 \up{0.74} \\ 
\midrule
Chain        & 63.40 \up{8.50} & 80.23 \up{1.38} & 66.12 \up{0.83} & 62.15 \up{5.14} & 88.57 \down{4.11} & 85.36 \up{0.47} & 74.31 \up{2.04} \\
Star         & 61.43 \up{6.53} & 82.34 \up{3.49} & 68.60 \up{3.31} & 64.95 \up{7.94} & 90.71 \down{1.97} & 82.74 \down{2.15} & 75.29 \up{3.02} \\
Random       & 62.75 \up{7.85} & 81.56 \up{2.71} & 69.42 \up{4.13} & 65.42 \up{8.41} & 90.00 \down{2.68} & \underline{87.29} \up{2.40} & 76.07 \up{3.80} \\
Complete     & 64.05 \up{9.15} & 80.91 \up{2.06} & 70.25 \up{4.96} & 62.15 \up{5.14} & 91.25 \down{1.43} & 83.13 \down{1.76} & 75.29 \up{3.02} \\
LLM-Debate   & 66.01 \up{11.11} & 81.35 \up{2.50} & \underline{71.07} \up{5.78} & \underline{65.89} \up{8.88} & 93.75 \up{1.07} & 87.19 \up{2.30} & 77.78 \up{5.51} \\ 
\midrule
AgentPrune   & 63.40 \up{8.50}  & 80.60 \up{1.75}  & 62.81 \down{2.48}  & 65.42 \up{8.41}  & 93.04 \up{0.36}  & 86.77 \up{1.88} & 75.34 \up{3.07} \\
AgentDropout & 64.05 \up{9.15}  & 82.41 \up{3.56}  & 62.81 \down{2.48}  & 61.68 \up{4.67}  & 93.21 \up{0.53}  & 86.10 \up{1.21} & 75.04 \up{2.77} \\
G-Designer   & 66.67 \up{11.77} & 81.27 \up{2.42}  & 66.12 \up{0.83}  & 64.49 \up{7.48}  & \underline{95.89} \up{3.21}  & 87.08 \up{2.19} & 76.92 \up{4.65} \\
ARG-Designer & \underline{67.32} \up{12.42} & \underline{83.87} \up{5.02}  & 68.60 \up{3.31}  & 63.55 \up{6.54}  & 93.57 \up{0.89}  & \underline{87.29} \up{2.40} & 77.37 \up{5.10} \\
\midrule
ATOM         & \textbf{72.55} \textbf{\up{17.65}} & \textbf{84.00} \textbf{\up{5.15}} & \textbf{71.90} \textbf{\up{6.61}} & \textbf{66.36} \textbf{\up{9.35}} & \textbf{96.61} \textbf{\up{3.93}} & \textbf{87.40} \textbf{\up{2.51}} & \textbf{79.80} \textbf{\up{7.53}} \\  
\bottomrule
\end{tabular}
}
\caption{Performance comparison (\%) on six benchmarks using Meta-Llama-3.1-8B-Instruct model. The best results are highlighted in bold, and the runners-up are underlined.}\label{tab:performance}
\end{table*}

\section{Experiments}

\subsection{Experimental Settings}

\textbf{Benchmarks:} We evaluate \ourmethod on six benchmarks spanning three domains: \textbf{General Reasoning} (MMLU~\citep{hendrycks2020measuring}), \textbf{Mathematical Reasoning} (GSM8K~\citep{cobbe2021training}, MultiArith~\citep{roy2016solving}, SVAMP~\citep{patel2021nlp}, AQuA~\citep{ling2017program}), and \textbf{Code Generation} (HumanEval~\citep{chen2021evaluating}). 

\textbf{Baselines:} We benchmark against a comprehensive suite of baselines across four paradigms: (1) \textbf{Single-agent methods}, including standard prompting and CoT~\citep{wei2022chain}; (2) \textbf{Static MAS topologies}, such as Chain, Star, Random, and Complete graphs~\citep{qian2024scaling}; (3) \textbf{Debate frameworks}, exemplified by LLM-Debate~\citep{du2023improving}; and (4) \textbf{Learnable MAS topologies}, including AgentPrune~\citep{zhang2024cut}, AgentDropout~\citep{wang2025agentdropout}, G-Designer~\citep{zhang2024g}, and ARG-Designer~\citep{li2025assemble}.

\textbf{Implementation Details.} Our backbone LLMs are Meta-Llama-3.1-8B-Instruct (served locally via vLLM) and DeepSeek-V3.2. We primarily conduct evaluations on the constrained 8B model to rigorously isolate our architectural gains from raw model capabilities. We use \texttt{all-MiniLM-L6-v2}~\citep{wang2020minilm} as encoder. Following standard MAS practice~\citep{zhuge2024gptswarm}, we default to 5 agents with explicit profiles.

\subsection{Experimental Results}

\begin{figure}[t]
    \centering
    \includegraphics[width=1\linewidth]{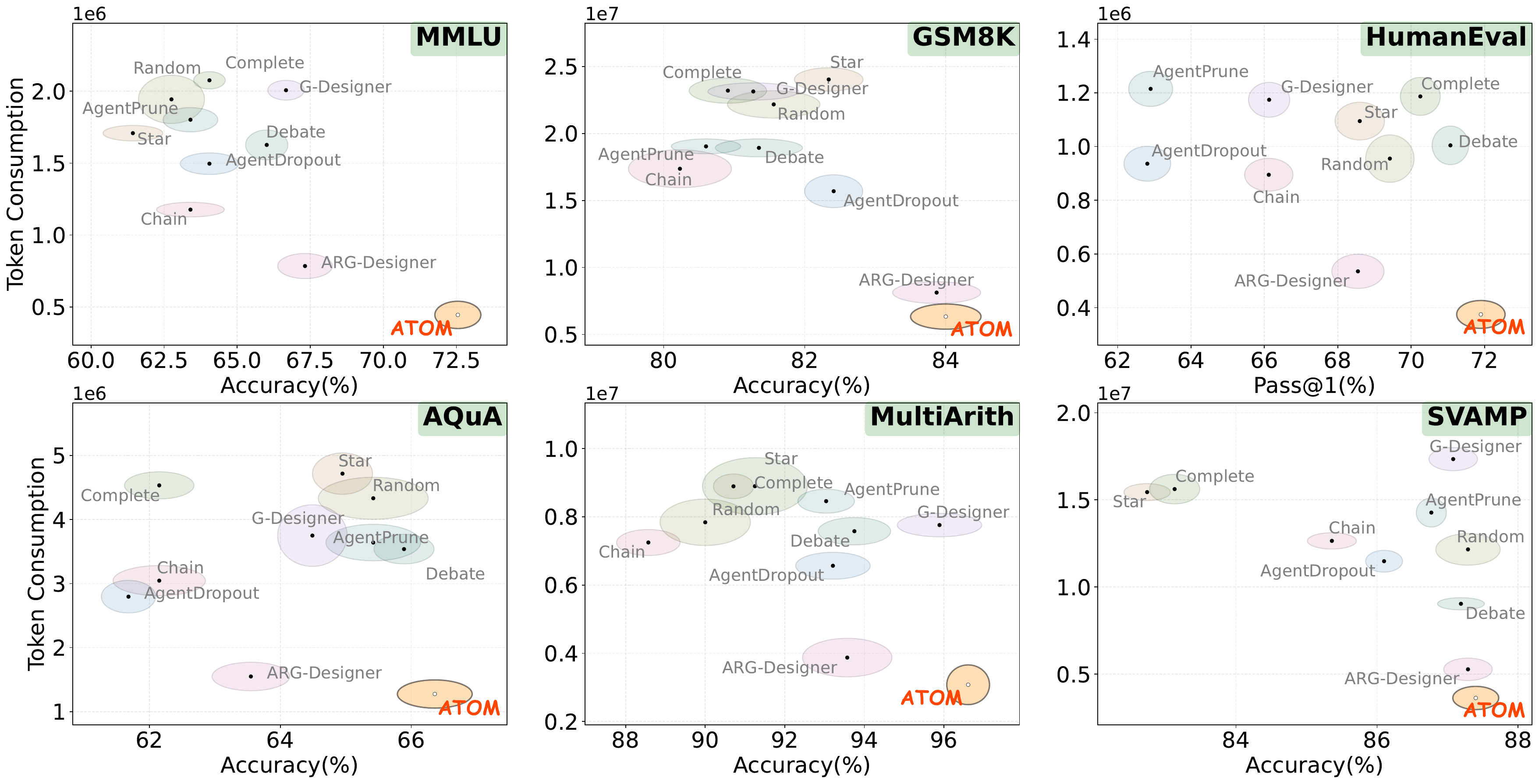}
    \caption{Comparison of performance and token consumption across baselines. Each point is an ellipse, whose height denotes the standard deviation of token consumption and width denotes the standard deviation of performance.}

    \label{fig:token}
\end{figure}

\subsubsection{Main Results}
The comparative results using the Meta-Llama-3.1-8B-Instruct model are summarized in Tab.~\ref{tab:performance}. \ourmethod achieves SOTA performance across all six benchmarks while drastically curtailing token consumption, establishing a strictly superior Pareto frontier (see Fig.~\ref{fig:token}).

\textbf{Superior Reasoning Accuracy.} \ourmethod consistently outperforms a broad range of baselines. Compared to the strongest learning-based baseline, ARG-Designer, it exhibits significant gains, notably achieving 72.55\% accuracy on MMLU (a 5.23\% absolute margin). Furthermore, it surpasses traditional methods like LLM-Debate by 2.86\% on MultiArith and 2.65\% on GSM8K, underscoring the advantage of adaptive topology over rigid, all-to-all communication protocols.

\textbf{Unmatched Token Efficiency.} A key advantage of \ourmethod is that it improves reasoning without incurring exponential token costs. Fig.~\ref{fig:token} shows that it minimizes prompt token (Ptok) overhead compared with MAS baselines. While methods such as AgentPrune and ARG-Designer suffer from communication bloat, \ourmethod keeps consumption close to the single-agent Vanilla baseline. On HumanEval, it achieves SOTA accuracy while reducing tokens consumption by over 30\% compared to ARG-Designer (from $\sim$5.3$\times$10$^5$ to 3.7$\times$10$^5$). By avoiding fixed structural costs and activating peripheral electrons only for complex queries, \ourmethod reduces redundancy while preserving capacity.

\subsubsection{Ablation Study}
To validate our core design choices, we conduct an ablation study on the complexity-aware budgeting strategy and architectural RL objectives (Table~\ref{tab:ablation}).

\textbf{Impact of Adaptive Budgeting.} We compare our adaptive strategy against Zero (nucleus only), Fixed, and Random budgets. Ignoring query difficulty via Fixed or Random budgets inflates token consumption and degrades performance (e.g., Fixed Budget accuracy drops to 63.64\% on HumanEval while nearly doubling MMLU prompt tokens from 0.29 to 0.51). Conversely, the highly token-efficient Zero Budget lacks topological capacity for complex tasks, causing a sharp 6.61\% accuracy drop on HumanEval. By explicitly modulating electron excitation based on estimated complexity, \ourmethod prevents resource waste on trivial tasks and under-provisioning on complex ones, achieving peak performance at near-minimal token cost.

\textbf{Impact of Architectural and RL Modules.} Replacing data-driven nucleus extraction with a random partition (\textit{w/o} Nucleus Iden.) causes a 3.92\% performance plunge on MMLU (from 72.55\% to 68.63\%) and a 4.96\% drop on HumanEval. This proves anchoring dynamic routing on a persistent cognitive backbone is vital for reliable reasoning within a sparse combinatorial space. Removing structural regularization (\textit{w/o} Str. Loss) triggers massive accuracy collapses (e.g., plunging to 64.05\% on MMLU and 62.81\% on HumanEval), confirming our center-heavy hierarchy prevents newly activated electrons from destabilizing the core conduit. Finally, excluding the sparsity constraint (\textit{w/o} Cost Penalty) inherently surges computational overhead---GSM8K prompt tokens jump from 3.88 to 5.23---while degrading HumanEval accuracy to 69.42\%, likely due to noisy context from redundant agents.

\begin{table*}[t]
\centering
\small
\resizebox{\linewidth}{!}{
\begin{tabular}{l|ccc|ccc|ccc} 
\toprule
\multirow{2}{*}{\textbf{Method}} & \multicolumn{3}{c|}{\textbf{MMLU}} & \multicolumn{3}{c|}{\textbf{GSM8K}} & \multicolumn{3}{c}{\textbf{HumanEval}} \\
 & \textbf{Perf.} & \textbf{Ptok.} & \textbf{Ctok.} & \textbf{Perf.} & \textbf{Ptok.} & \textbf{Ctok.} & \textbf{Perf.} & \textbf{Ptok.} & \textbf{Ctok.} \\        
\midrule
\textbf{\ourmethod} & \textbf{72.55} & \textbf{0.29} & \textbf{0.12} & \textbf{84.00} & \textbf{3.88} & \textbf{1.72} & \textbf{71.90} & \textbf{0.26} & \textbf{0.11} \\
\midrule
Zero Budget   & 68.60 & 0.29 & 0.13 & 80.22 & 4.21 & 1.08 & 65.29 & 0.14 & 0.06 \\
Fixed Budget  & 69.28 & 0.51 & 0.20 & 79.83 & 4.98 & 1.50 & 63.64 & 0.16 & 0.07 \\
Random Budget & 67.97 & 0.48 & 0.20 & 81.63 & 5.27 & 1.29 & 67.77 & 0.48 & 0.20 \\
\midrule
\textit{w/o} Nucleus Iden. & 68.63 & 0.31 & 0.13 & 80.45 & 4.02  & 1.47 & 66.94 & 0.21 & 0.09 \\
\textit{w/o} Str. Loss     & 64.05 & 0.33 & 0.13 & 80.68 & 4.98  & 1.50 & 62.81 & 0.07 & 0.03 \\
\textit{w/o} Cost Penalty  & 69.28 & 0.32 & 0.14 & 82.23 & 5.23  & 1.98 & 69.42 & 0.27 & 0.12 \\
\bottomrule
\end{tabular}
}
\caption{Ablation study on key components of \ourmethod.}
\label{tab:ablation}
\end{table*}

\subsubsection{Extended Analysis}
\label{sec:extended_analysis}

\begin{figure}[b] 
    \centering
    
    \begin{minipage}[b]{0.68\linewidth}
        \centering
        \includegraphics[width=1\linewidth]{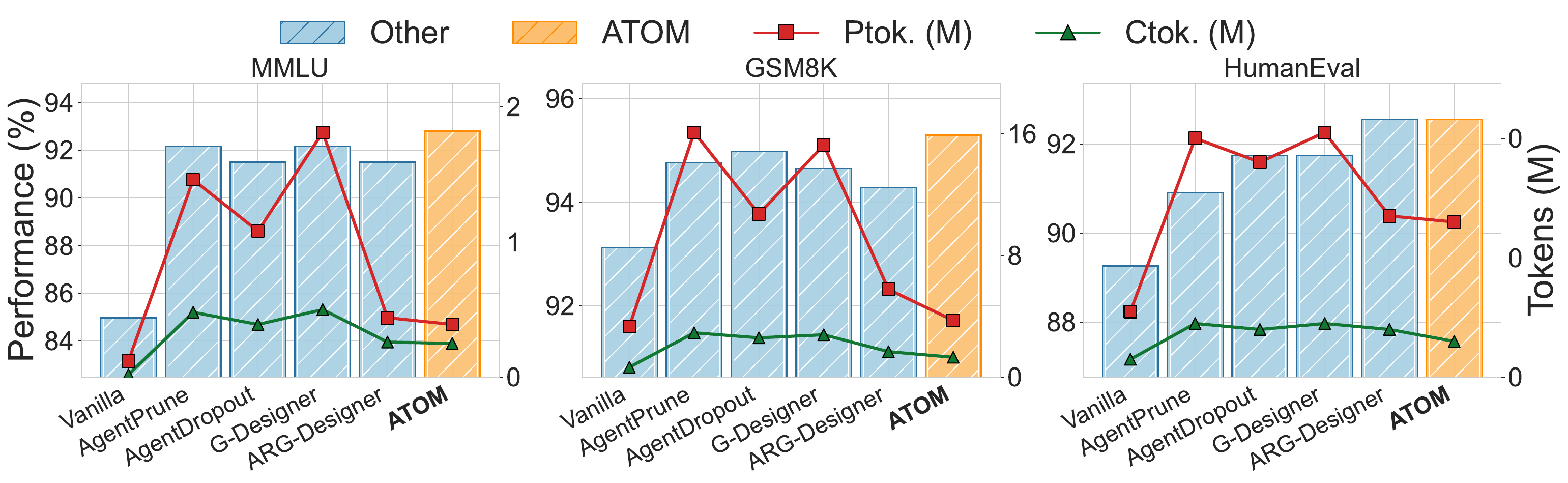}
    \end{minipage}
    \hfill
    \begin{minipage}[b]{0.28\linewidth}
        \centering
        \includegraphics[width=1\linewidth]{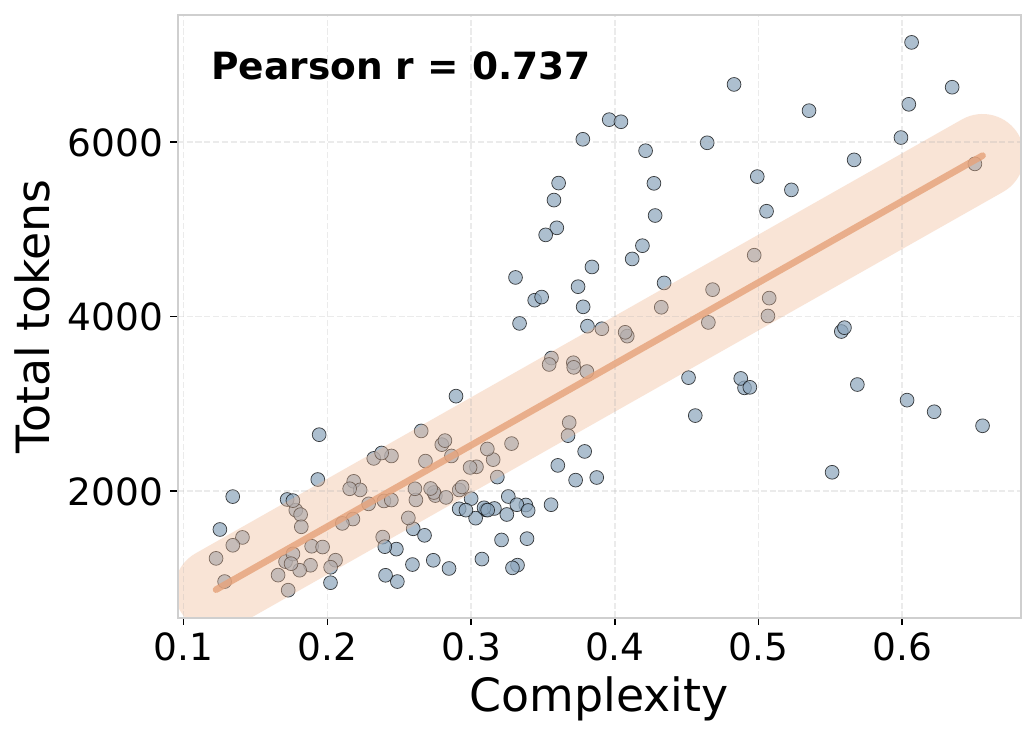}
    \end{minipage}
    
    \begin{minipage}[t]{0.68\linewidth}
        \caption{Performance and token consumption comparison using DeepSeek-V3.2. \textbf{Ptok.} and \textbf{Ctok.} denote the number of prompt and completion tokens, respectively, measured in millions (M, \(10^6\)).}
        \label{fig:performance-deepseek}
    \end{minipage}
    \hfill
    \begin{minipage}[t]{0.28\linewidth}
        \caption{Token consumption with respect to MMLU complexity.}
        \label{fig:mmlu-complexity-tokens}
    \end{minipage}
    
\end{figure}
\textbf{Scaling to Advanced Backbones.} 
To test whether framework-level gains persist with stronger backbones, we evaluate \ourmethod on DeepSeek-V3.2. As shown in Fig.~\ref{fig:performance-deepseek}, \ourmethod maintains highest accuracy across all datasets while incurring computational overhead close to the Vanilla single-agent baseline. By contrast, alternative MAS frameworks such as AgentPrune and G-Designer require up to $4\times$ more tokens. These results demonstrate that \ourmethod remains a cost-effective and scalable topology generation framework even for frontier large language models.

\textbf{Efficacy of Adaptive Budgeting.} Predicted task complexity and actual token consumption on MMLU show a strong positive correlation (Pearson $r = 0.737$, Fig.~\ref{fig:mmlu-complexity-tokens}). This confirms \ourmethod dynamically aligns its topological scale with intrinsic difficulty. The minimal nucleus efficiently resolves simple queries, while constraint-heavy tasks unlock the peripheral electron reservoir, demonstrating that our theoretical budgeting mechanism translates directly into proportional computational scaling.

\begin{figure}[t!]
    \centering
    \begin{minipage}[b]{0.48\linewidth}
        \centering
        \includegraphics[width=1\linewidth]{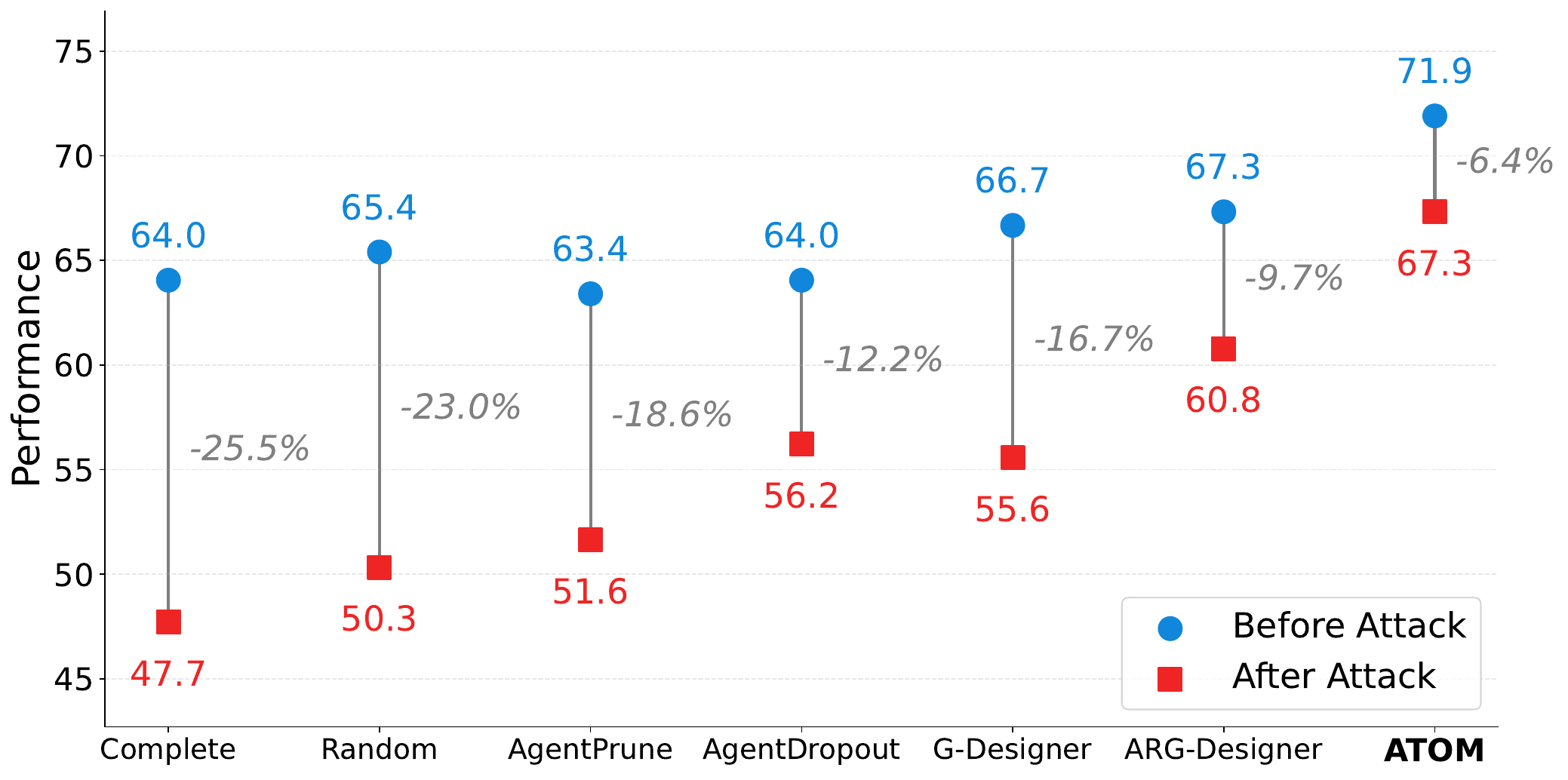}
        \caption{Robustness analysis}
        \label{fig:attack}
    \end{minipage}
    \hfill
    \begin{minipage}[b]{0.48\linewidth}
        \centering
        \includegraphics[width=1\linewidth]{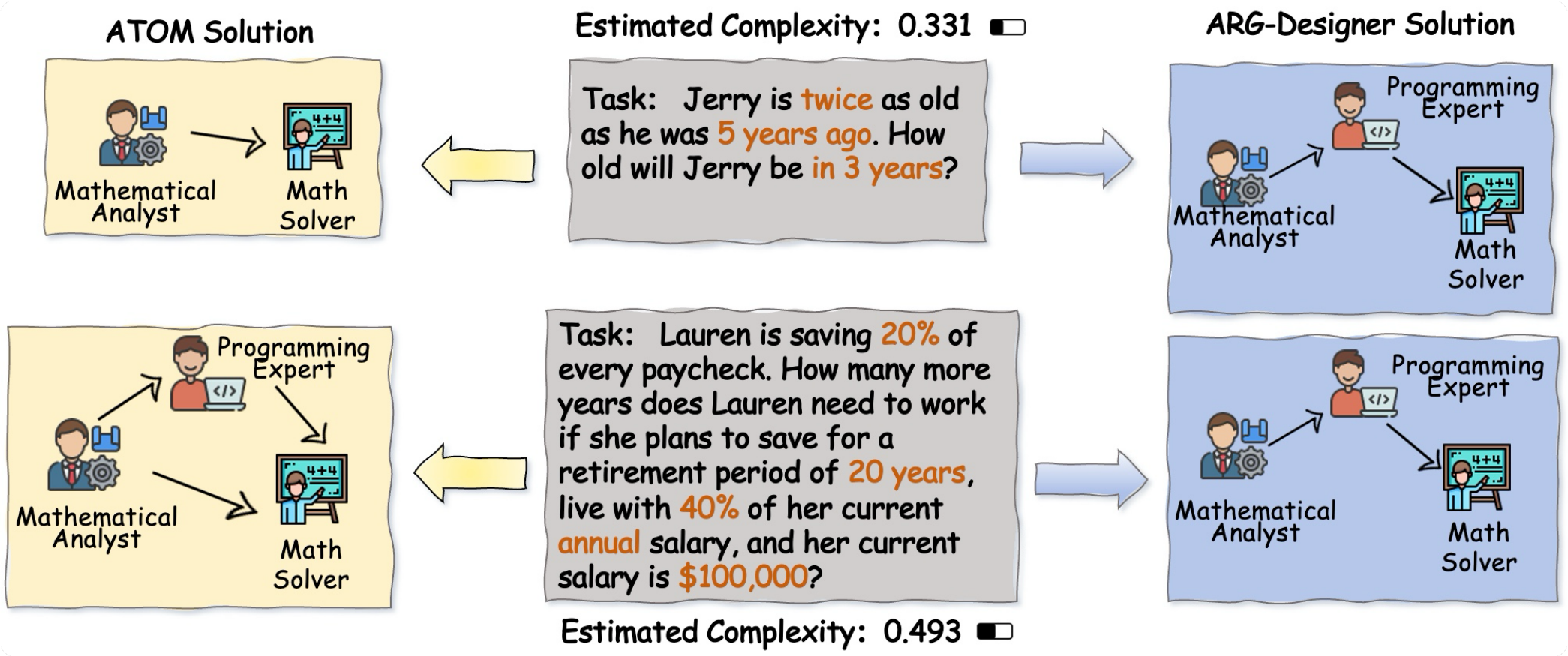}
        \caption{Cases of ATOM and ARG-Designer}
        \label{fig:case}
    \end{minipage}
\end{figure}

\textbf{Robustness Against Prompt Injection.} Following~\citep{zhuge2024gptswarm}, we evaluate resilience against prompt injection. While adversarial instructions catastrophically degrade fully dynamic MAS topologies, \ourmethod exhibits superior robustness, limiting the performance drop to 6.4\% (Fig.~\ref{fig:attack}). This stems from our two-tier hierarchy and structural regularization. By anchoring execution to the offline-verified nucleus, compromised electrons are structurally prevented from eclipsing the primary cognitive conduit, thereby mitigating cascading systemic failures.

\subsubsection{Case Study}
To illustrate \ourmethod's structural adaptability, Fig.~\ref{fig:case} compares its generated topologies against ARG-Designer~\citep{li2025assemble} on GSM8K queries. Crucially, \ourmethod avoids the ``average-complexity'' trap by explicitly aligning topological scale with task demands. For a simple arithmetic task (estimated complexity $0.331$), it instantiates a minimal two-agent chain (Mathematical Analyst $\rightarrow$ Math Solver) to prevent redundant overhead. Conversely, for a multi-step financial problem (complexity $0.493$), it dynamically expands the reasoning network by recruiting a \emph{Programming Expert}. In stark contrast, ARG-Designer exhibits structural rigidity, generating the exact same three-agent topology regardless of query difficulty, thereby validating \ourmethod's dual advantage in efficiency and flexible scalability.

\section{Related Work}

\paragraph{LLM-based Multi-Agent System}
While LLM-based agents excel across diverse domains~\citep{liu2024large,zhong2024debug,dong2024self,xie2024waitgpt,li2024dawn,hong2025data,zhao2025video,yue2025survey,zhu2024autotqa}, single agents inherently struggle with long-horizon tasks due to limited reasoning depth and expertise coverage~\citep{du2023improving,valmeekam2023planning,ke2025survey}. To address these limitations, Multi-Agent Systems (MAS) leverage collective intelligence~\citep{islam2024mapcoder,zhang2025sortinghat} to achieve superior performance in complex, collaborative scenarios such as software development and world simulations~\citep{qian2023chatdev,hong2023metagpt,chen2024scalable,park2023generative,kaiya2023lyfe}.

\paragraph{MAS Topology Optimization}
Topology design has evolved from static structures---like chains~\citep{wei2022chain, hong2023metagpt} or trees~\citep{yao2023tree}---to adaptive frameworks~\citep{chen2023autoagents}. Recent deployment optimization efforts span three categories: structural pruning and dropout (e.g., AgentPrune~\citep{zhang2024cut}, AgentDropout~\citep{wang2025agentdropout}), budget-constrained initialization~\citep{cai2025agentbalance,yang2025bamas,tian2025agentinit}, and from-scratch GNN-based generative design (e.g., G-Designer~\citep{zhang2024g}, ARG-DESIGNER~\citep{li2025assemble}). 
Additionally, emerging works tackle communication costs by modeling bandwidth as a scarce resource via auction-based bidding to filter low-value messages~\citep{fan2026cost}. Concurrently, GoAgent~\citep{chen2026goagent} reduces overhead by forming explicit task-specific groups rather than relying on dense node-centric connectivity.
Our work departs fundamentally from both template modification and purely generative prediction. Inspired by flexible-firm architectures~\citep{volberda1999building}, \ourmethod\ introduces a hierarchical strategy, structurally decoupling a stable reasoning core from dynamic, budget-aware expansion.

\section{Conclusion}
In this paper, we propose \ourmethod, a reinforcement learning framework with a stable nucleus backbone and dynamic electron extensions. By decoupling offline structure learning from online query-conditioned expansion, \ourmethod adaptively allocates computation and instantiates task-specific topologies. Experiments on six benchmarks show that \ourmethod achieves state-of-the-art performance with up to 30\% higher token efficiency and strong scalability. 

\bibliographystyle{ref}  
\small
\bibliography{custom}
\normalsize

\end{document}